\documentclass[5p,times]{elsarticle}
\usepackage{graphicx}
\usepackage{textcomp}
\usepackage{dcolumn}
\usepackage{bm}
\usepackage{amsmath}
\usepackage{multirow}

\newcommand{\mean}[1]{\mbox{$\langle{#1}\rangle$}}

\usepackage{makecell}
\usepackage[printwatermark]{xwatermark}
\usepackage[abspath]{currfile}
\usepackage[para,online,flushleft]{threeparttable}

\usepackage[colorlinks=true]{hyperref}
\usepackage{lineno}
\usepackage[colorinlistoftodos,prependcaption,textsize=tiny]{todonotes}

\begin{document}
\begin{frontmatter}
\title{Photoinjector Generation of High-Charge Magnetized Beams \\
for Electron-Cooling Applications}
\author[1]{A. Fetterman}
\author[1]{D. Mihalcea}
\author[2]{S. Benson}
\author[3]{D. Crawford}
\author[3]{D. Edstrom}
\author[2]{F. Hannon}
\author[1,3,4]{P. Piot}
\author[3]{J. Ruan}
\author[2]{S. Wang}
\address[1]{Department of Physics and Northern Illinois Center for Accelerators \& Detectors Development, Northern Illinois University, DeKalb IL 60115, USA} 
\address[2]{Thomas Jefferson National Laboratory, Newport News VA 23660, USA} 
\address[3]{Fermi National Accelerator Laboratory, Batavia, IL 60510, USA}
\address[4]{Argonne National Laboratory, Lemont, IL 60510, USA}
\begin{abstract}
Electron-cooling offers a relatively simple scheme to enable high-luminosity collisions in future electron-ion and hadron colliders. Contemplated TeV-energy hadron colliders require relativistic (sub 100 MeV) high-charge [${\cal O}(\mbox{nC})$] electron beams with a specific transverse eigenemittance partition. This paper discusses the generation of high-charge ($Q\le 3.2$~nC) 40 MeV electron bunches with eigenemittance partition consistent with requirements associated with electron-cooling option for future electron-ion colliders. The supporting experiment was performed at the FAST facility at FermiLab. The data are compared with numerical simulations and the results discussed in the context of beam requirement for future electron-ion colliders. 
\end{abstract}


\end{frontmatter}

\section{Introduction}
High-intensity beams in storage rings have been employed as probes in elementary-particle physics. There is an ultimate limit on the attainable brightness of circulating beams. For hadrons, the brightness limit comes from beam-beam and intra-beam effects, which gradually increase the emittance and energy spread of the beam, thus reducing the brightness. Consequently, devising efficient and fast phase-space cooling is critical for future hadron and electron-ion colliders. Over the years, there have been several cooling techniques categorized as ($i$)\ stochastic and ($ii$) frictional methods. Stochastic processes consist of detecting information associated with the circulating-bunch distribution to apply a corrective kick. Repeating such action over many turns eventually reduces the beam's phase-space volume. Examples of stochastic methods include microwave and optical stochastic cooling~\cite{RevModPhys.57.689,PhysRevLett.71.4146,PhysRevE.50.3087}, and micro-bunched electron cooling~\cite{PhysRevLett.111.084802}. Frictional techniques include laser-based radiative cooling~\cite{PhysRevLett.61.826} and electron cooling~\cite{Budker:1967sd}. 

In electron cooling, a ``cold" electron beam co-propagates with the ion beam. In the ion-beam rest frame, the electrons (which have the same average velocity) produce an equivalent friction force. Since its experimental demonstration fifty years ago~\cite{Budker:1021068}, electron cooling has been implemented in several proton and ion storage rings~\cite{Nature1990}. In most configurations, the electron and hadron beams interact within a long solenoid providing a uniform axial magnetic field $B_s$. 
In addition to guiding the electron beam, the magnetic field can be combined with an angular-momentum-dominated -- or magnetized -- electron beam to possibly enhance the cooling efficiency~\cite{Derbenev:1978qd,Derbenev:1978}. Given the conservation of magnetic flux, the electrons do not have any ``coherent" motion but instead all electrons cycle on small helices with a radius given by the gyroradius. Producing, accelerating, and transporting relativistic ($\gamma \ge 50$) magnetized beams $-$ i.e. beams with large kinetic angular momentum $-$ with parameters consistent with future electron-ion collider is a challenging problem. So far the formation of magnetized beams in photoinjectors has focused on low-charge 15-MeV beam~\cite{PhysRevSTAB.7.123501}, or on producing a beam with large canonical angular momentum (CAM) for conversion to a flat beam with extreme transverse-emittance ratios~\cite{Xu:2019xjg}.

The present paper discusses the experimental generation of 1.6 and 3.2~nC with eigenemittance partitions satisfying requirements for a potential magnetized electron cooling option in an electron-ion collider.
\section{Principle of magnetized-beam generation}
The production of a magnetized beams from an electron source is staged. First a beam with significant CAM is formed. Then the CAM is converted into kinetic angular momentum. Producing a CAM-dominated beam consists in subjecting the cathode surface to an axial magnetic field $B_c$ resulting in an electron with radial position $r$ acquiring an ab-initio angular momentum $P_{\theta}=eA_{\theta}=\frac{eB_c}{2}r +{\cal O}(r^3)$ under the paraxial approximation and neglecting the stochastic momentum arising from the emission process. By averaging the associated angular momentum over $L\equiv rP_{\theta}=\frac{eB_c}{2}r^2$, we conveniently characterize the magnitude of the CAM contribution to the beam dynamics via the magnetization~\cite{PhysRevSTAB.6.104002}
\begin{eqnarray}
{\cal L} \equiv \frac{\langle L \rangle}{2mc} = \frac{eB_c}{2mc}\sigma_c^2, 
 \end{eqnarray} 
where $\sigma_c\equiv\langle r^2 \rangle^{1/2}/2$ is the rms transverse size (assumed to be identical along the two transverse directions). In practical units the magnetization is given by ${\cal L}\mbox{[\textmu{m}]}  \simeq 294 B_c\mbox{[T]}(\sigma_c\mbox{[mm]})^2$. 
The magnetization plays a  role similar to the emittance in the ensemble-averaged beam-envelope equation and it is customary to introduce the effective emittance as~\cite{Lee:1107845,Nagaitsev:439908}
\begin{eqnarray}~\label{eq:effemit}
 \varepsilon_{\mbox{\tiny eff}} = [{\cal L}^2 +  \varepsilon_{u}^2]^{1/2}, 
 \label{effective_emit}
\end{eqnarray} 
where $\varepsilon_{u}$ represents the uncorrelated (or intrinsic) emittance arising from the emission process and possible nonlinear effects (e.g. space charge forces). Given the cross-plane coupling, the projected emittances are not invariants of motion and one instead resorts to introducing eigenemittance obtained from the $4\times4$ beam covariance matrix $\Sigma$ in the  $(x,\frac{p_x}{mc},y,\frac{p_y}{mc})$ phase space by solving the characteristic equation 
\begin{eqnarray}
\mbox{det}[J_4 \Sigma - i\varepsilon_{\ell} I_4]=0.
\end{eqnarray}
Here $J_4\equiv \begin{pmatrix} J  & 0 \\ 0 & J \end{pmatrix}$ $-$ with $J\equiv \begin{pmatrix} 0  & -1 \\ 1 & 0 \end{pmatrix}$ $-$ and $I_4$ is the $4\times 4$ identity, $\ell\in [1,4]$ indexes the eigenvalues, and $i\equiv\sqrt{-1}$. The latter equation is derived from the facts that ($i$) $J_4 \Sigma$ transforms accordingly to a similarity transformation under the action of a beamline transfer matrix and ($ii$) its eigenvalues are shown to be imaginary~\cite{PhysRevLett.64.1073}. In practice the eigenvalue are degenerate so that we introduce the $\varepsilon_{\pm}$ emittances (with the  convention $\varepsilon_+>\varepsilon_-$) as  $\varepsilon_1=\varepsilon_2\equiv \varepsilon_+$, and $\varepsilon_3=\varepsilon_4\equiv \varepsilon_-$. These eigenemittances are related to the effective emittance via 
\begin{eqnarray}~\label{eq:eigenemit}
\varepsilon_{\pm}= \varepsilon_{\mbox{\tiny eff}} \pm {\cal L},
\label{eq:eigen-emitt0}
\end{eqnarray}
Considering a CAM-dominated beam, i.e. with ${\cal L} \gg \varepsilon_{u}$, the associated eigenemittances simplify to  
 \begin{eqnarray}
\varepsilon_{+}=  2{\cal L}\equiv \varepsilon_{d}, \mbox{and~} \varepsilon_{-}=\frac{ \varepsilon_{u}^2}{2 {\cal L}}\equiv \varepsilon_{c},
\label{eq:eigen-emitt}
\end{eqnarray}
where $\varepsilon_{d}$ and $\varepsilon_{c}$ are respectively referred to as the normalized ``drift" and ``cyclotron" emittances~\cite{Derbenev2000AdvancedOC,PhysRevSTAB.3.094002,PhysRevE.66.016503}. 
The importance of eigenemittance partition to the cooling process stems from the interaction in the cooling solenoid channel. As the CAM-dominated beam enters the cooling solenoid-channel section, its RMS size is set to  $\sigma_s=\left(\frac{Bc}{Bs}\right)^{1/2}\sigma_c$, where $B_s$ is the peak axial magnetic field in the cooling solenoid. Such a matching condition ensures the collective rotation experienced by the beam inside the solenoid cancels and the electrons follow a tight helical trajectory along the B-field lines thereby enhancing the electron-ion interaction. \\

\begin{figure}[hthhh!!!]
\centering
\includegraphics[width=0.45\textwidth]{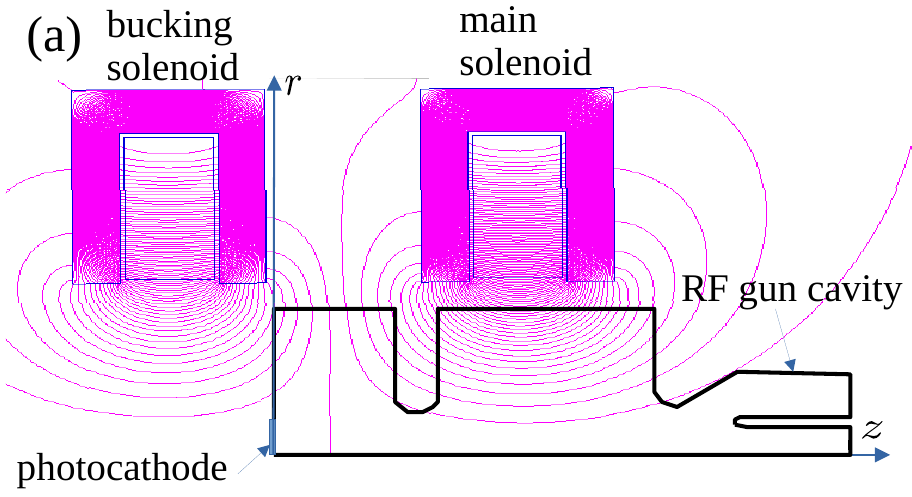}
\includegraphics[width=0.48\textwidth]{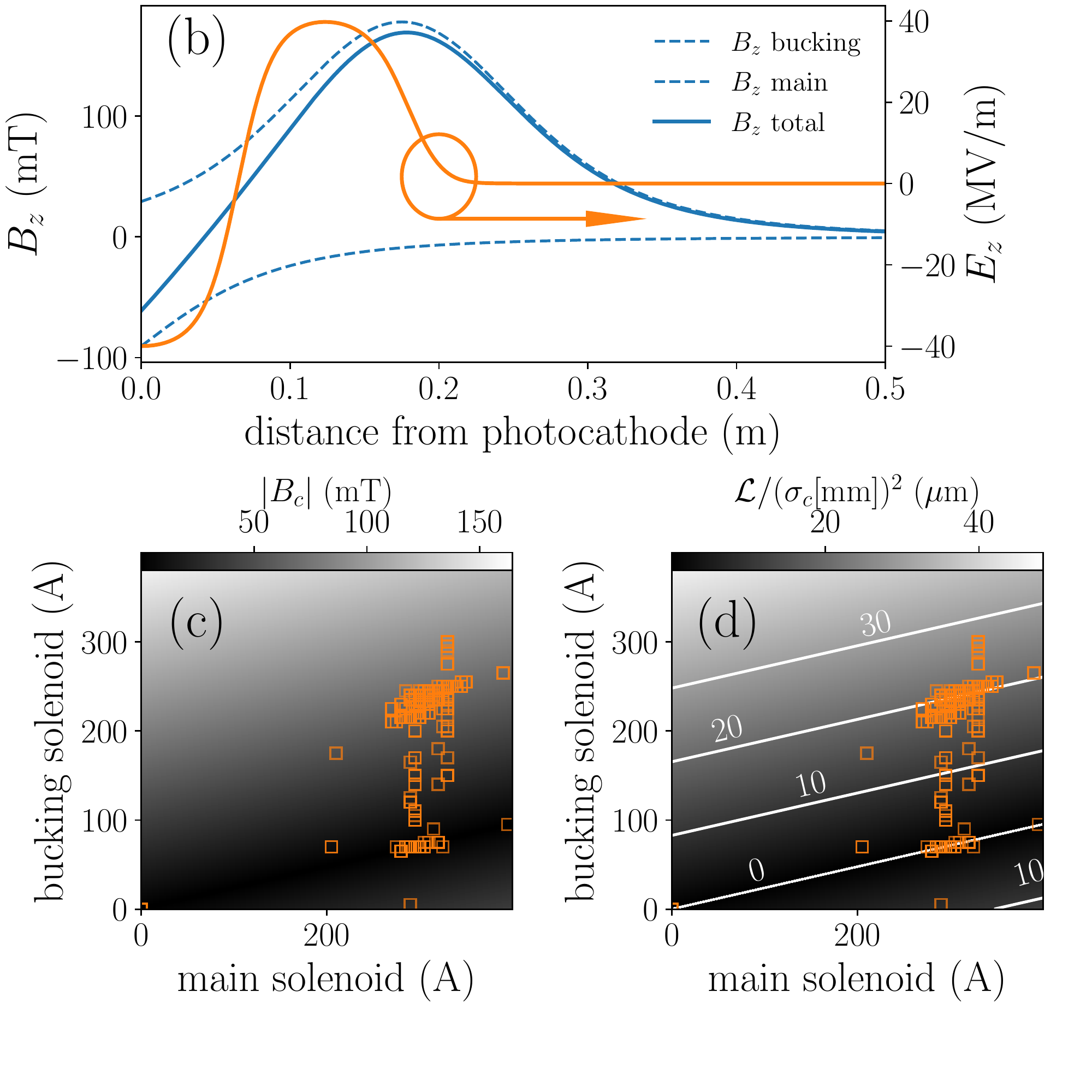}
\caption{Configuration of the FAST photoemission electron source with magnetic field line (magenta lines) and RF gun cavity (a) and corresponding axial magnetic (blue trace) and electric (orange trace) fields in the RF gun region (b). Dependence of the axial magnetic field on the photocathode surface $B_c$ (c) and corresponding  normalized magnetization per unit beam size (d) on main- and bucking-solenoid excitation currents. For these calculation the bucking polarity is set to its nominal value (i.e. opposite to the main solenoid). The orange symbols in plots (c) and (d) represent experimental values explored during the experiments reported in this paper.}
\label{fig:magnetization}
\end{figure}

The formation of magnetized beams in a normal-conducting radio-frequency (RF) photoinjector is straightforward and consists in immersing the photocathode in a magnetic field~\cite{PhysRevSTAB.7.123501}. A diagram of the photoinjector used for the experiment reported in the next sections appears in Fig.~\ref{fig:magnetization}(a) along with an example of field distribution throughout the RF photoinjector in Fig.~\ref{fig:magnetization}(b). In brief, the photoinjector consists of a $1+\frac{1}{2}$-cell L-band RF gun ($f=1.3$~GHz) nested in a pair of solenoid lenses (referred to as the bucking and main solenoids). Under typical non-magnetized beam operation, the bucking solenoid cancels the magnetic field on the photocathode. The attainable magnetic field on the cathode and associated magnetization are summarized in  Fig.~\ref{fig:magnetization}(c, d) as a function of possible bucking and main solenoid settings.

The generation of magnetized beam discussed in this paper was motivated by a potential colliding-energy cooling scheme for the Jefferson Lab Electron-Ion Collider (JLEIC) design~\cite{Zhang:2019afy} initially proposed as one of the electron-ion collider design options. However, the results apply to a possible implementation of magnetized electron cooling in the Electron-Ion Collider (EIC) complex currently in construction~\cite{Montag:2019rpr}. Table~\ref{tab:jleic} compares the required electron-beam parameters of the JLEIC project with those attainable at FAST (either experimentally verified or simulated). The FAST facility~\cite{Antipov_2017} can in principle produce the needed transverse beam parameters but with a much shorter bunch length. Also, the peak-to-peak fractional energy spread is substantially higher at FAST due to the effect of RF-induced curvature on the longitudinal phase space while the uncorrelated fractional energy spread is comparable to the one required for the JLEIC electron cooling. 

\begin{table}[ht]
\caption{Comparison of electron-cooling beam requirements for JLEIC with the ones achievable (inferred from simulations) at FAST. The JLEIC parameters are adapted from Ref.~\cite{Benson:2018jzt}. All the values are RMS quantities and the emittance values are normalized. \label{tab:jleic} }
\begin{center}
\begin{threeparttable}
\begin{tabular}{l c c c  c c}\hline\hline\
                  &        & JLEIC       &       FAST \\
                  &        & strong cooling &  \\
\hline
parameter  & unit & value & value  \\
\hline
beam energy    &  MeV & [20,55] &  $\le 45$~\tnote{a} \\
bunch charge   & nC            & 3.2 (1.6) & $>3.2$~\tnote{a} \\
cathode spot size\footnote{JLEIC requirements give the cathode radius $r_c$ so that RMS values are taken to be $\sigma_{x,y}=r_c/2$, i.e. assuming a {\em uniform} emission source}  & mm            & 1.55  & 1 \\
$B$ field on cathode       & T            & 0.05 &  $<0.09$~\tnote{a} \\
cyclotron emit.  & \textmu{m}  & $\le 19$  &  $<5$ \\
drift  emit.         & \textmu{m}  & 36   & 37  \\
$\delta p/p$  (slice)  & $-$      &     $3\times 10^{-4}$ &   $<4\times 10^{-4}$\\
$\delta p/p$  (pk-to-pk.)  & $-$  & $<6\times 10^{-4}$   &  ${\cal O}(10^{-2})$~\tnote{b} \\
bunch length $\sigma_z$   &  cm      &     2  &   $\sim{0.25}$ \\
\hline
\hline
\end{tabular}
\begin{tablenotes}
\item[a] Values experimentally achieved. \item[b] This value includes correlated energy spread.
\end{tablenotes}

\end{threeparttable}

\end{center}
\end{table}
\begin{figure}[ht]
\centering
\includegraphics[width=0.48\textwidth]{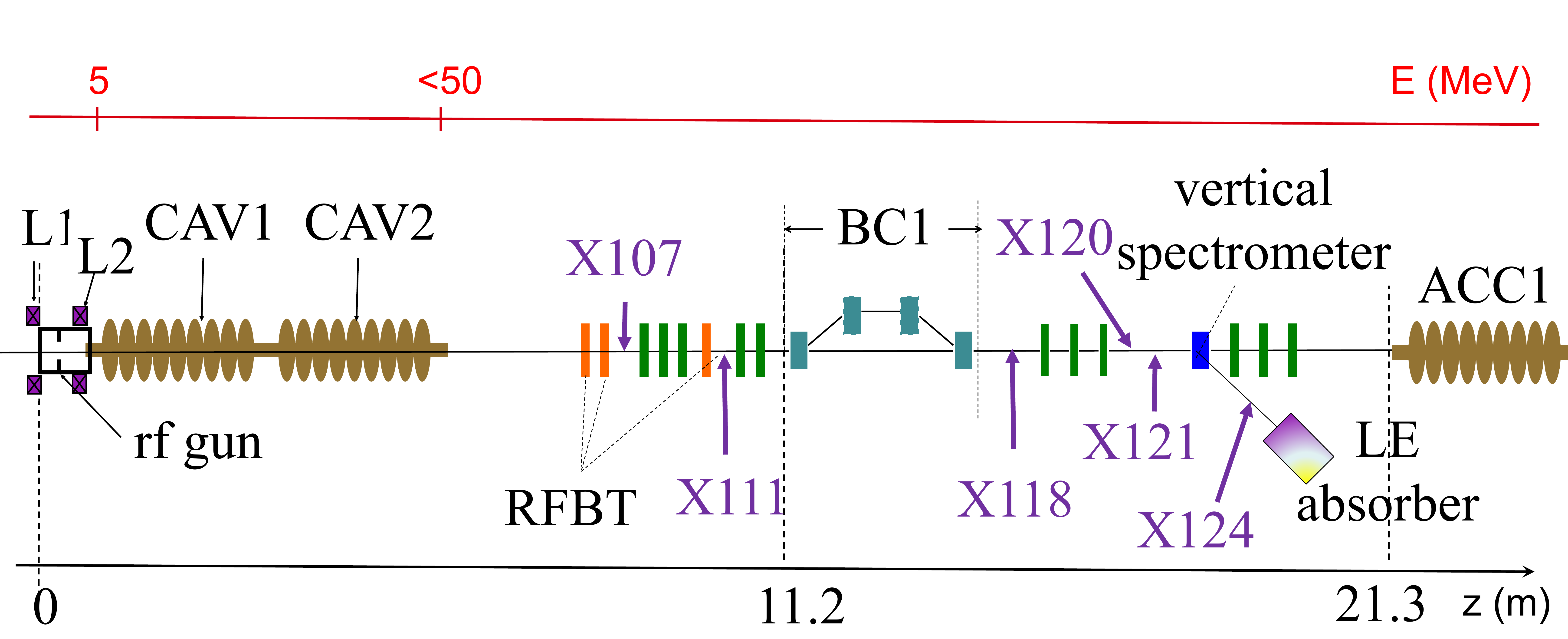}
\caption{Schematics of the FAST injector low energy section. The "X" labels refer to diagnostics stations with OTR and/or YAG screens. Diagnostics stations X107 and X118 are also equipped with horizontal and vertical multi-slit masks for emittance measurements. }
\label{fig:fast_schematics}
\end{figure}

\section{Optimization of the FAST injector via particle tracking}

The FAST photoinjector~\cite{Piot:2010zz} employed in the experiment is diagrammed in Fig.~\ref{fig:fast_schematics}. In brief, the beam is photoemitted from a Cs$_2$Te photocathode located in a $1\frac{1}{2}$ cell RF gun; see Fig.~\ref{fig:magnetization}(a). The  $\sim{4}$ MeV beam is then accelerated in a pair of TESLA-type superconducting RF (SRF) cavities (CAV1 and CAV2). Downstream of the accelerating section the beamline includes a suite of beam diagnostics (labeled as "X"), quadrupole magnets, and a bunch compressor (BC1) to manipulate the beam before further acceleration in an 8-cavity accelerating cryomodule (ACC1). The experiment reported in this paper was performed in the injector area upstream of the accelerating cryomodule. Before the experiment, the injector operating parameters were optimized using {\sc astra}~\cite{Klaus:2018} to guide possible experimental configurations. The optimizations were performed for the two cases of charges considered (1.6 and 3.2~nC) using a multi-objective algorithm. The laser spots size on the photocathode and peak field of the bucking and main solenoids were variables in the optimization process. The other beamline (cavity phase and field) and photocathode-laser settings were set to values corresponding to their nominal operating point. The two main objectives were to produce a magnetized beam with normalized drift emittance around $\varepsilon_{d} \simeq 36$~\textmu{m} while minimizing the four-dimensional normalized emittance defined as $\varepsilon_{4D}\equiv (\varepsilon_{d}\varepsilon_{c})^{1/2}$. 

Table~\ref{tab:param} summarizes an example of optimized and selected beamline parameters that fulfill JLEIC electron-cooling emittance specifications for 1.6 and 3.2~nC.

\begin{figure}[ht]
\centering
\includegraphics[width=0.48\textwidth]{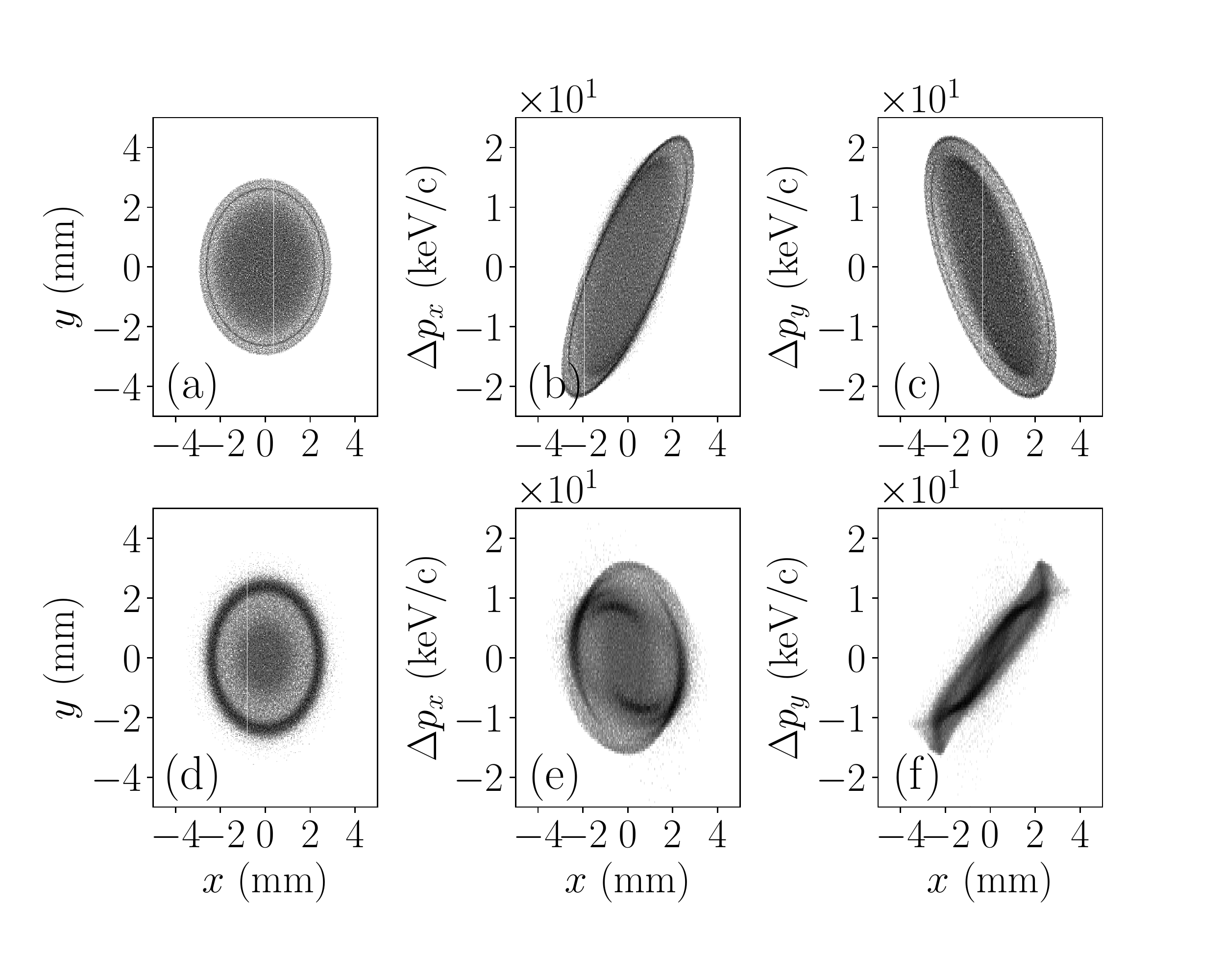}
\caption{Comparison of simulated beam transverse distribution (a,d), horizontal phase space (b,e) and $x-p_y$ distribution (c,f) for bunch charges of 1.6 (upper row) and 3.2~nC (lower row). The distributions are recorded at $z=8$~m from the photocathode. }
\label{fig:astraTPS}
\end{figure}

\begin{figure}[ht]
\centering
\includegraphics[width=0.48\textwidth]{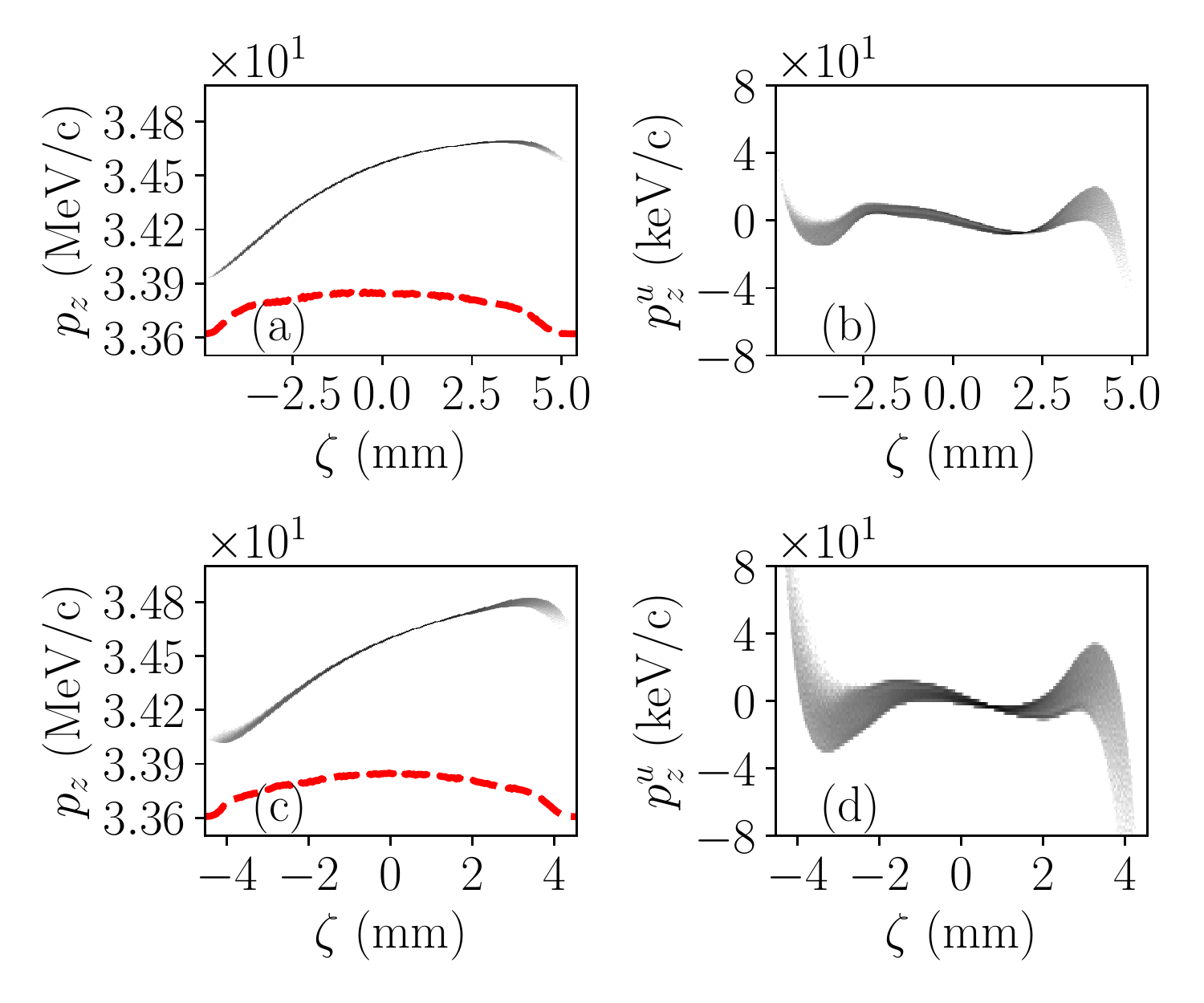}
\caption{Comparison of longitudinal phase space (left column, the red dash trace represents the current projection along the longitudinal coordinate), and uncorrelated longitudinal phase space (LPS, right column)  simulated with {\sc astra} for 1.6 (a,b) and 3.2~nC (c,d) bunches. The distributions are taken at $z=8$~m from the photocathode. For plots (b,d) the uncorrelated momentum is defined as $p_z^u=p_z-\sum_{i=1}^5 \mu_i \zeta$ where the $\mu_i$ coefficients are obtained from a polynomial regression of the LPS. The head of the bunch corresponds to $\zeta>0$.}
\label{fig:astraVSimpacttLPS}
\end{figure}

\begin{table}[hbt]
\caption{Optimized accelerator settings for magnetized-beam generation at FAST with final emittances consistent with JLEIC requirements; see Fig.~\ref{fig:astraVSimpacttLPS}. The optimization was performed for two cases of bunch charges (1.6 and 3.2~nC). \label{tab:param} }
\begin{center}
\begin{threeparttable}
\begin{tabular}{l c c c c}\hline\hline\
parameter  & symbol & \multicolumn{2}{c}{value} & unit \\
\hline
bunch charge   & $Q$  & 1.6   & 3.2 & nC \\
laser rms duration   & $\sigma_t$ &  \multicolumn{2}{c}{3}  & ps     \\
laser rms spot size  & $\sigma_c$ &  0.75 & 1.52  & mm     \\
$B$ field on cathode & $B_c$  & 118 & 28 & mT  \\
bucking sol. peak B field & $B_b$ & 67 &  0.0 & mT \\ 
main sol. peak B field    & $B_m$ & 174 & 171 & mT \\ 
laser launch phase   & $\varphi_g$  &  0~\tnote{a}  & 0~\tnote{a} & deg \\
E field on cathode  & $E_g$  &  \multicolumn{2}{c}{40} &  MV/m  \\
CAV1 phase  & $\varphi_1$  &  \multicolumn{2}{c}{0~\tnote{a}}   &   deg \\
CAV1 peak E field  & $E_1$  &  \multicolumn{2}{c}{28} &   MV/m  \\
CAV2 phase  & $\varphi_2$  &  \multicolumn{2}{c}{0~\tnote{a}}   &   deg \\
CAV2 peak E field  & $E_2$  &  \multicolumn{2}{c}{28} &  MV/m  \\
\hline
\hline
\end{tabular}
\begin{tablenotes}
\item[a] All phases are referenced w.r.t. the maximum-energy-gain phase.
\end{tablenotes}

\end{threeparttable}
\end{center}

\end{table}
\begin{figure}[ht]
\centering
\includegraphics[width=0.46\textwidth]{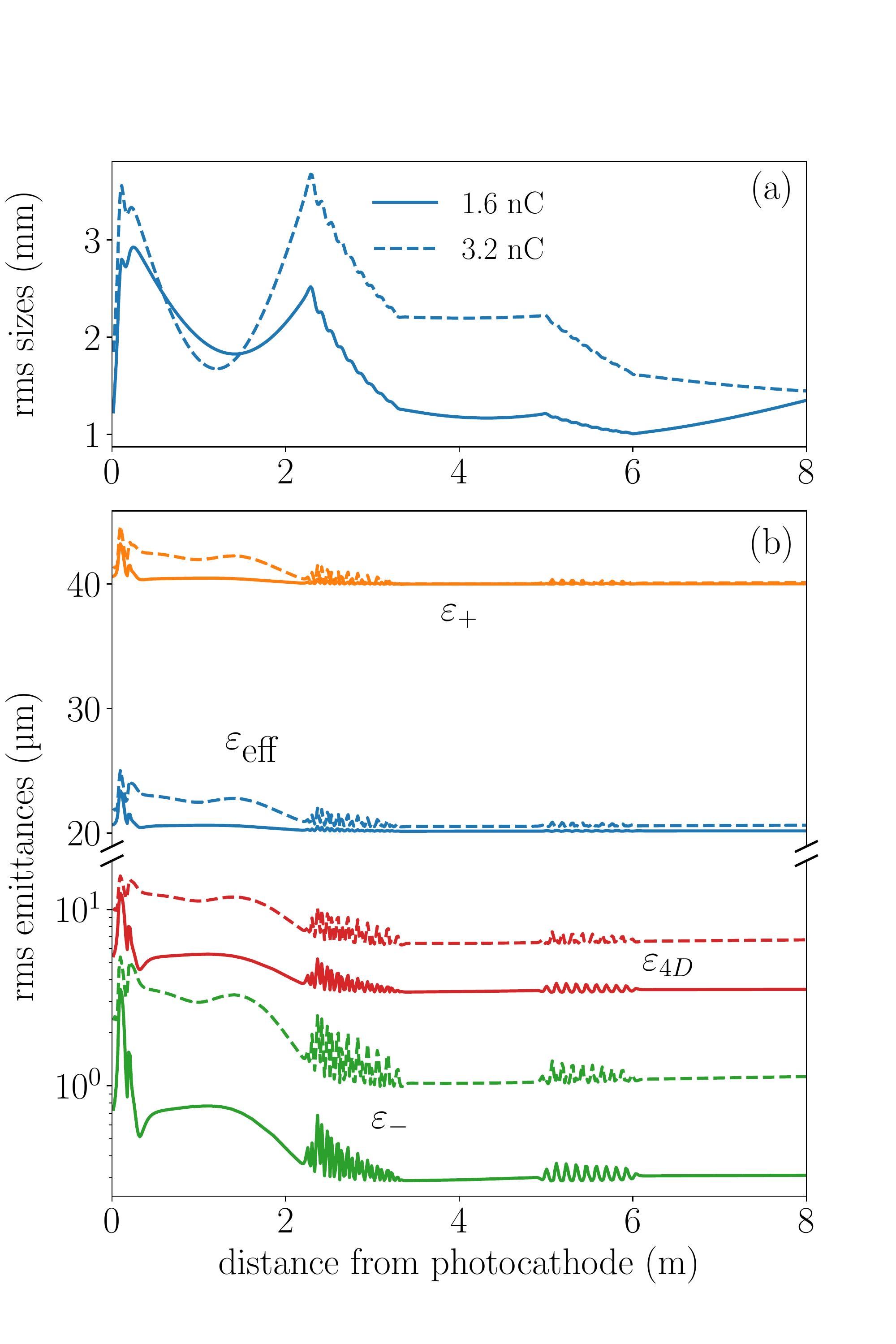}
\caption{Simulated transverse rms beam sizes (a) and normalized  emittances (b) for 1.6- (solid traces) and 3.2-nC (dashed traces) bunches. The emittances include the eigenemittances $\varepsilon_{\pm}$, the effective emittance $\varepsilon_{\mbox{\tiny eff}}$, and the 4D emittance $\varepsilon_{4D}$. }
\label{fig:emitevol}
\end{figure}
The evolution of the normalized emittances along the photoinjector beamline appears in Fig.~\ref{fig:emitevol}. For both the low and high charge cases, similar effective and eigenemittance values $\varepsilon_{ +}$ are achieved, while the 1.6-nC bunch has a $\sim 2$-times smaller 4D-emittance value, resulting in $\varepsilon_{-}$ a factor 4 smaller than the one associated with the 3.2-nC bunch. The transverse phase-space distributions computed at $z=8$~m from the photocathode appear in Fig.~\ref{fig:astraTPS} using 200k macroparticles. Similar transverse beam distributions are produced while the phase-space and cross-plane correlations are different owing to the difference in focusing along the beamline [see Fig.~\ref{fig:emitevol}(a)] and the charge-dependent space-charge effects. The numerical simulations also indicate that the beam dynamics downstream of the CAV2 SRF cavity is dominated by the CAM contribution. Additionally, it worth noting that the simulations indicate the presence of peripheral ring-like structures in the beam transverse distribution $(x,y)$ which becomes more pronounced at 3.2~nC and is accompanied with some tail formation especially visible in Fig.~\ref{fig:astraTPS}(d-f). These features were reproduced with another beam-dynamics program and are associated with a phase-space bifurcation which could ultimately result in halo generation. Finally, the longitudinal phase-space distributions associated with the 1.6 and 3.2-nC bunches appear in Fig.~\ref{fig:astraVSimpacttLPS}(a,b). Given the rather long bunch, the energy spread is dominated by nonlinear correlation impressed by the RF curvature during acceleration in the cavities CAV1 and CAV2 while the intrinsic momentum spread is at the keV/c level as seen in Fig.~\ref{fig:astraVSimpacttLPS}(b,d). 
%
%
%
%
%
\section{Measurement of the magnetized beam emittances and relation to eigenemittances}

\subsection{Experimental method overview}
One method to infer the eigenemittance is to transform the incoming magnetized beam into a flat beam with a round-to-flat beam adapter as discussed in Ref.~\cite{PhysRevSTAB.4.053501,PhysRevSTAB.6.104002,PhysRevSTAB.9.031001}. The method, which requires the precise tuning of skew-quadrupole magnets, was evaluated and eventually forwent as it is challenging to implement. Additionally, the mapping of the eigenemittances to projected emittances is generally non-ideal due to the significant incoming energy spread (which leads to chromatic effects) and space-charge effects. Instead, a simpler technique based on measuring the magnetized beam to recover the eigenemittances was employed. The method is adapted from the work presented in Ref.~\cite{PhysRevAccelBeams.22.102801}. Using a scanning slit to intercept the magnetized beam and subsequently analyzing the transmitted beamlets enables the measurement of the effective $\varepsilon_{\mbox{\tiny eff}}$ and uncorrelated $\varepsilon_u$ emittances which can then be directly related to the eigenemittance following Eqs.~\ref{eq:eigenemit} and \ref{eq:effemit} as
\begin{eqnarray}
\varepsilon_{n,\pm}= \varepsilon_{n,\mbox{\tiny eff}} 
    \pm \sqrt{\varepsilon_{n,\mbox{\tiny eff}}^2-\varepsilon_{n,u}^2}.
\label{eq:eigen-emitt_meas}
\end{eqnarray}
In our experimental setup, the magnetized round beam is intercepted by either one of two scanning slits, one vertical and the other one horizontal, located $\sim 16$~meters downstream of the photocathode (and labeled as X118 in Fig.~\ref{fig:fast_schematics}). The transverse distribution of the transmitted beamlet is recorded on a YAG screen (shown as X121 in Fig.~\ref{fig:fast_schematics}) positioned 1.78~m downstream of the slit.
\begin{figure}[ht]
  \centering
  \includegraphics[width=0.45\textwidth]{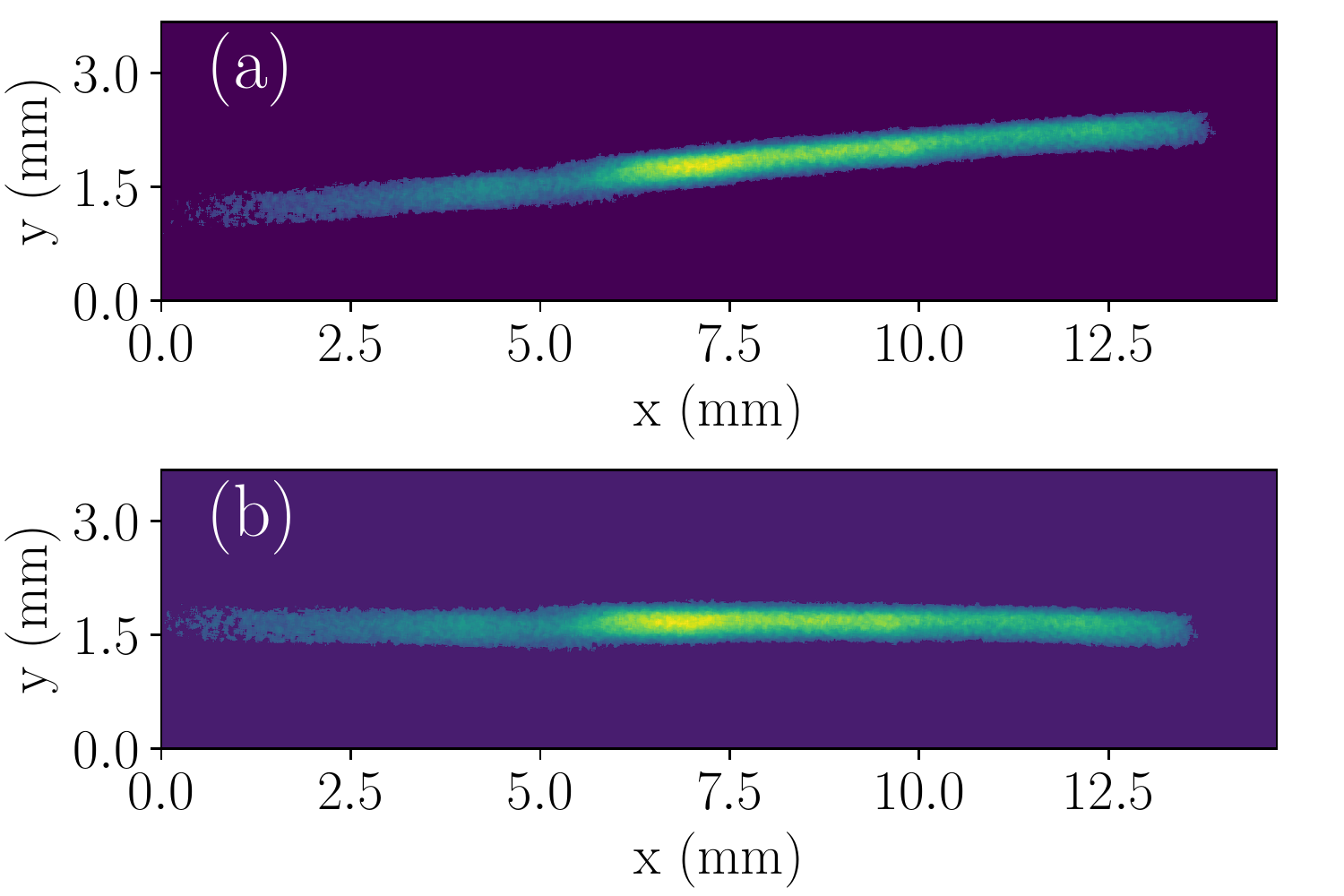}
  \caption{Image of a rotated beamlet when magnetization is ${\cal L} \simeq 26$~\textmu{m} and beam charge is 3.2~nC (a). The beamlet image is rotated to subtract the contribution of the beam angular momentum (b).}
  \label{slit}
\end{figure}  
The beamlet transmitted through the slit undergoes a shearing before reaching the YAG screen due to the angular momentum inherited from the magnetized round beam. The distance between the slit and the YAG screen is sufficient to allow for the beamlet to be tilted by several degrees on the screen; see Fig.~\ref{slit}(a). The corresponding beamlet size [vertical size in the case of Fig.~\ref{slit}(b)] is directly related to its divergence as an electron with position $x_i$ at the slit has a final position $X_i=x_i+Lx'_i$ at the YAG screen (here $x'_i$ is the electron divergence at the slit and $L\simeq 1.78$~m the distance between the slit and YAG screen). Consequently, the rms divergence of the beamlet can be computed as  
\begin{eqnarray}
\sigma_{x'}=\left[\sigma_X^2- \frac{w^2}{12}\right]^{1/2},
\label{xp}
\end{eqnarray}
where we took $\mean{x^2}\simeq w^2/12$ with $w\simeq 40$~\textmu{m} being the slit width. Consequently, the horizontal emittance can be retrieved following well-established measurement techniques~\cite{lejeune-1980}. The same process can be repeated for the vertical emittance with the substitution $x \leftrightarrow y$. Nominally, for the case of a magnetized beam, the divergence is the convolution of two contributions: the beamlet shearing related to the beam magnetization, and an uncorrelated angular spread associated with the uncorrelated emittance.  Nominally, the emittance measured corresponds to the effective emittance $\varepsilon_{\mbox{\tiny eff}}$. However, the contribution from the beam magnetization can be removed during the analysis by rotating the beamlet image such that the $x-y$ correlation averaged over all the beamlets is removed; see Fig.~\ref{slit}(b). The corresponding beamlet size [vertical size in the case of Fig.~\ref{slit}(b)] provides information on the uncorrelated divergence and enables the measurement of the uncorrelated emittance $\varepsilon_u$. 

Consequently, our analysis consists of measuring both the correlated and uncorrelated emittance in both the horizontal and vertical planes to retrieve the eigenemittances following Eq.~\ref{eq:eigen-emitt_meas}. For our measurements, we record the beamlets for 20 slits positions across the incoming beam. For all our measurements the laser spot on the cathode was held constant and its rms horizontal and vertical sizes were measured to be $\sigma_{c,x}=0.90$~mm and $\sigma_{c,y}=1.21$~mm respectively. 

\subsection{Magnetization}

 A scan was performed for each beam magnetization. The beam magnetization was varied by changing the magnitude of the magnetic field at photocathode surface using the currents in the bucking and main solenoids. There were 35 scans corresponding to magnetization values ${\cal L}  \in [ 0, \sim 30]$~\textmu{m}. 

\begin{figure}[hthhh!!!!!!]
  \centering
  \includegraphics[width=0.465\textwidth]{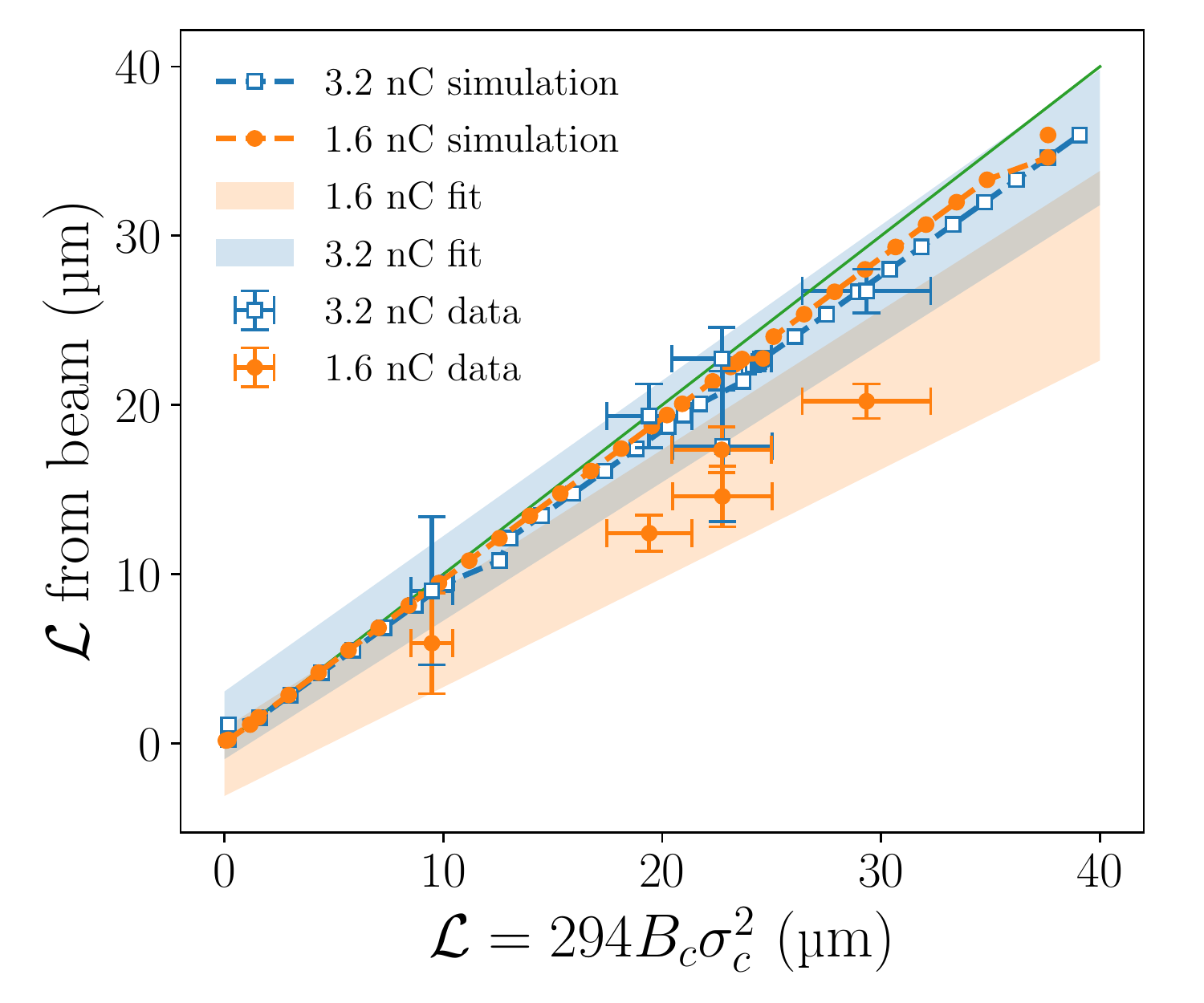}
  \caption{Measured beam magnetization as a function of the initial magnetization computed from solenoid settings and photocathode-laser spot size. The data symbols with errorbars are compared with simulation (symbols with dash lines). The solid green diagonal is drawn simply to aid the eye. The shaded areas represent error on the fit within 98\% confidence level.}
  \label{mag}
\end{figure}

Figure~\ref{mag} compares the simulated and measured magnetization. In simulations the magnetization is evaluated by averaging the angular momentum at the slit position over the $N$ macroparticles composing the beam as ${\cal L}  =\frac{1}{mc N}\sum_{i=1}^N [x_i p_{y,i} -y_i p_{x,i}]$. The simulations use the experimental set points for all systems and the initial macroparticle distributions were generated via Monte-Carlo sampling of the two-dimensional laser distribution recorded on a ``virtual" cathode diagnostics providing a real-time one-to-one optical image of the cathode surface. The magnetization computed from the magnetic field takes the geometric average of the cathode spot sizes $\sigma_c=\sqrt{\sigma_{c,x}\sigma_{c,y}}$ and consider the magnetic field obtained from the current setpoint based on magnetic measurement of the solenoid. Ideally, one would expect the magnetization measured from the beam to equate the magnetization inferred from the magnetic field on the photocathode. The deviations of simulated values from the ideal scaling are approximately 7.7\% for the 3.2~nC beam and 4.3\% for the 1.6~nC beam. The charge dependence of the simulated magnetization suggests that it is only slightly affected by beam collective effects. The values of magnetization derived from beam effective and uncorrelated emittances (see Eqn.~\ref{effective_emit}) agree well with expectations only for the 3.2~nC case. The rotation angles were underestimated for the 1.6~nC case possibly due to an alignment problem during data collection. Therefore, for the following analysis, we subsequently report the magnetization values evaluated from solenoid settings and measured photocathode-laser spot sizes. 

\subsection{Eigenemittances}

\begin{figure}[ht]
  \centering
  \includegraphics[width=0.48\textwidth]{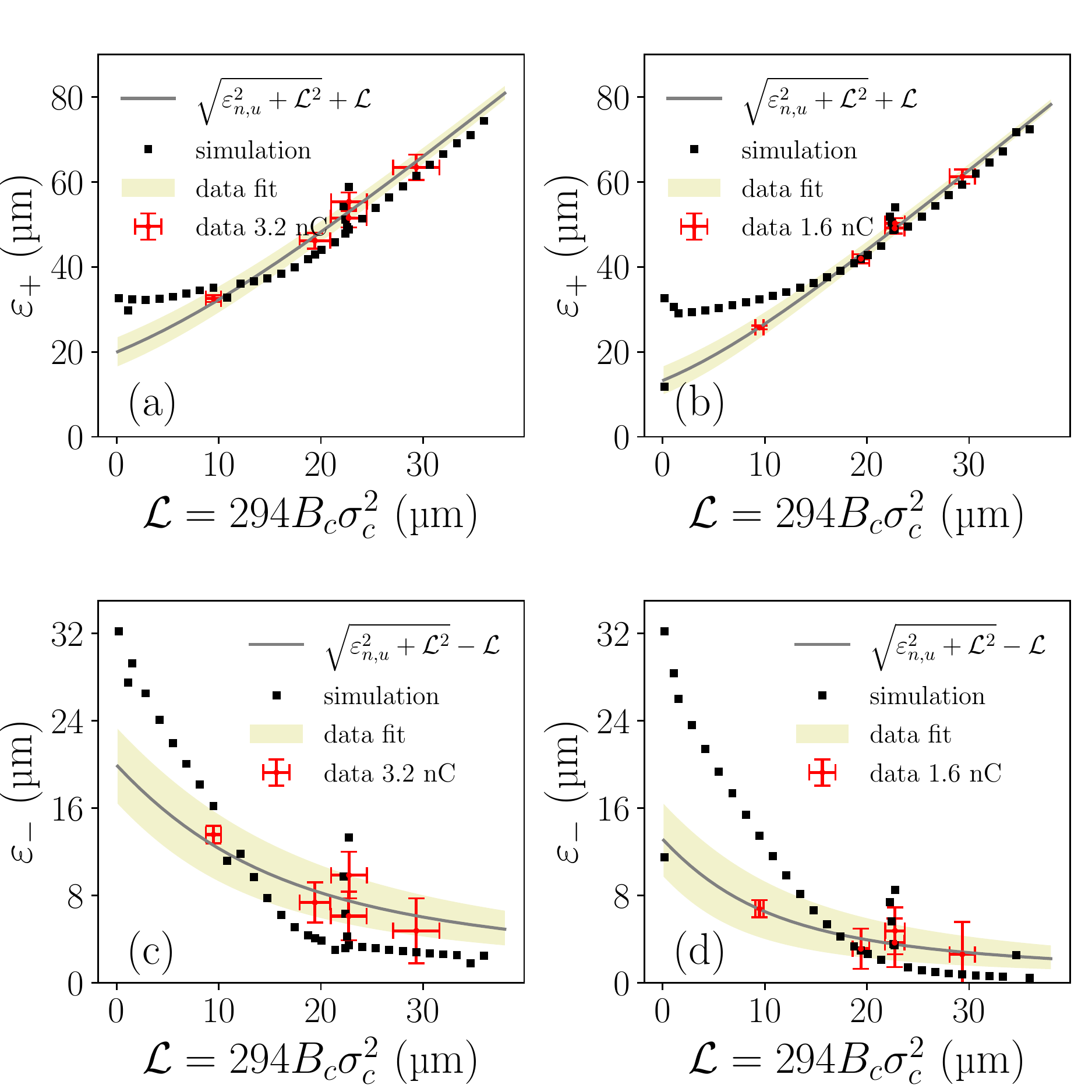}
  \caption{Measured and simulated eigenemittances: $\varepsilon_{+}$ for 3.2~nC (a), $\varepsilon_{+}$ for 1.6~nC (b), $\varepsilon_{-}$ for 3.2~nC (c) and $\varepsilon_{-}$ for 1.6~nC (d). All graphs include a fit of the measured data with the theoretical expressions of the eigenemittances. The fit parameter is allowed within $\pm 3\sigma$ to ensure a 98~\% confidence level (CL).}
  \label{ee_sim_data}
\end{figure}

Figure~\ref{ee_sim_data} summarizes the measured and simulated values of the eigenemittances. The plots include a fit of the data points with Eq.~\ref{eq:eigen-emitt_meas}. The fit parameter, in this case the uncorrelated emittance, is allowed to vary within $\pm 3\sigma$ from the central value reported by the fitting routine. This ensures that the extrapolated phase space from the fit (shaded areas in the plots) has 98\% confidence level (CL). The values of the magnetization and eigenemittances
are also summarized in Table~\ref{tab:eigen-emitt}.    

To determine the transverse beam size, the relative beamlet intensities and positions of the whole beam centroid relative to the slit are needed. The positions are derived from beam position monitors (BPMs) readings and the uncertainties on the beamlet position were determined to be dominated by statistical fluctuations. These errors range from 50 to 100~\textmu{m} and include BPMs resolution of about 40~\textmu{m}. Likewise, the angular spreads obtained in the two cases are used to evaluate the effective and uncorrelated emittances. The errors of the angular spread include statistical fluctuations and experimental uncertainties and were found to be 0.2 to 0.3~mrad for the effective emittance and 0.05 to 0.1~mrad for the uncorrelated emittance.

\begin{table}[ht]
\caption{Measured normalized eigenemittances for 1.6~nC and 3.2~nC bunch charges.\label{tab:eigen-emitt} }
\begin{center}
\begin{tabular}{l  c c  c c}\hline\hline\
                 &  \multicolumn{2}{c}{1.6~nC} &
                 \multicolumn{2}{c}{3.2~nC} \\
\hline
$\mathcal{L} $ (\textmu{m}) & $\varepsilon_{ +}$ (\textmu{m}) & $\varepsilon_{-}$ (\textmu{m}) & $\varepsilon_{ +}$ (\textmu{m}) & $\varepsilon_{-}$ (\textmu{m}) \\ 
\hline
9.5 & $25.7 \pm 0.5$ & $6.8 \pm 0.5$ & $32.5 \pm 0.8$ & $13.6 \pm 0.8$ \\
19.4 & $41.9 \pm 1.1$ & $3.1 \pm 1.1$ & $46.2 \pm 1.8$ & $7.4 \pm 1.8$ \\
22.7 & $49.1 \pm 1.3$ & $3.7 \pm 1.3$ & $51.5 \pm 2.2$ & $6.1 \pm 2.2$ \\
22.8 & $50.3 \pm 1.2$ & $4.8 \pm 1.2$ & $55.4 \pm 2.1$ & $9.9 \pm 2.1$ \\
29.3 & $61.3 \pm 1.7$ & $2.6 \pm 1.7$ & $63.4 \pm 3.0$ & $4.5 \pm 3.0$ \\
\hline
\hline
\end{tabular}
\end{center}
\end{table}

At low levels of magnetization ($\mathcal{L}  < 5$~\textmu{m}) the beam becomes emittance-dominated and the extrapolation of the eigenemittance fit in Fig.~\ref{ee_sim_data} is unreliable. 

As expected, the uncorrelated emittance is not uniquely determined by the magnetization. In Fig.~\ref{ee_sim_data} such a fact can be observed for magnetizations around $\sim 23$~\textmu{m} where there is a cluster of simulation and experimental data. A given magnetization value can be obtained with different combinations of solenoid currents which impact the emittance-compensation process. Ultimately, producing the brightest beam for the desired magnetization requires one to select the solenoid settings that minimize the uncorrelated emittance. Such an approach was followed for the multi-objective optimization presented in Fig.~\ref{fig:emitevol} where the objectives were to produce a given value for $\varepsilon_+$ while minimizing the four-dimensional emittance $\varepsilon_{4D} =\varepsilon_{u}$.

\subsection{Energy spread}
\begin{figure}[ht]
  \centering
  \includegraphics[width=0.45\textwidth]{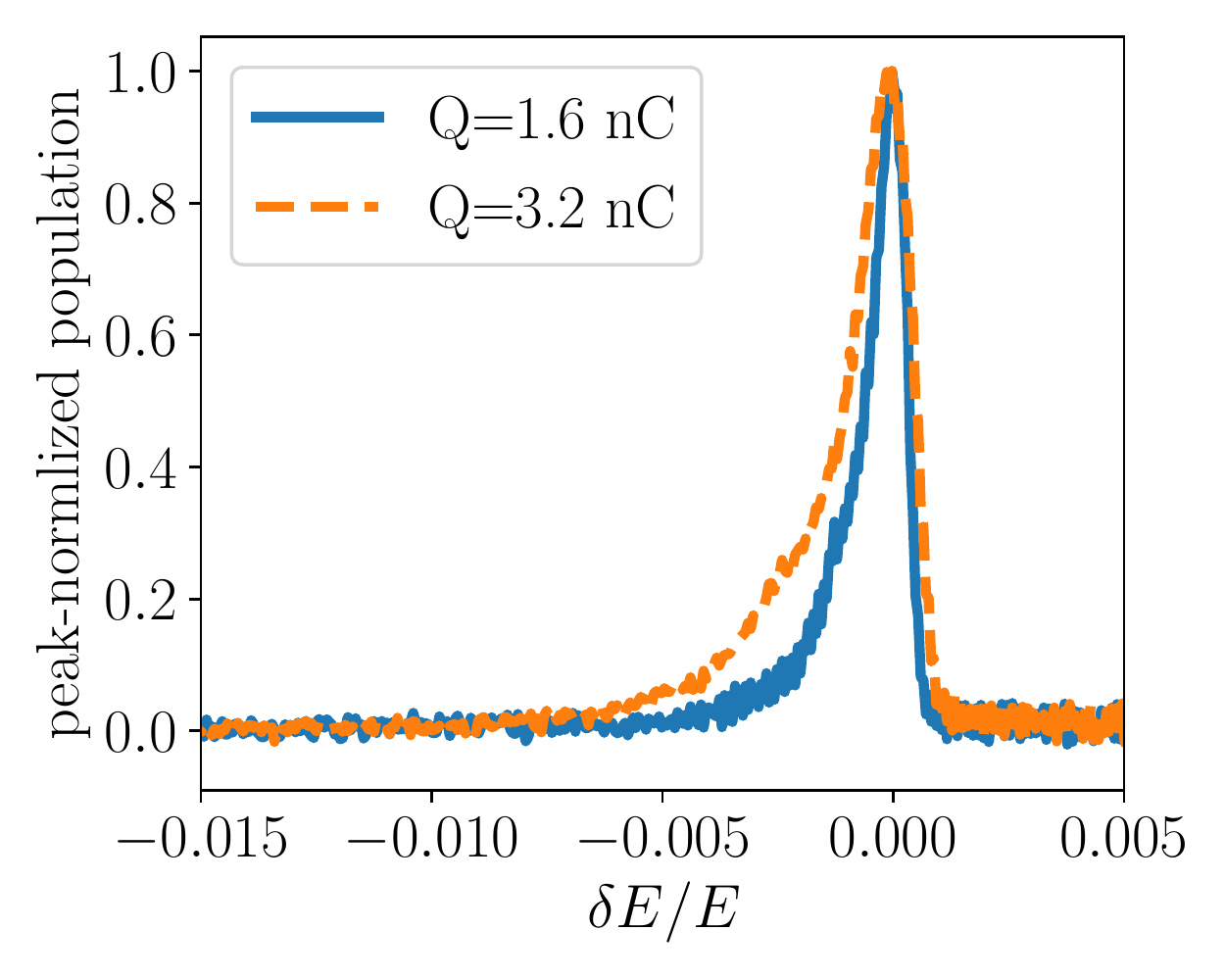}
  \caption{Fractional energy spread distribution for the 1.6 (solid blue trace) and 3.2~nC (dash orange trace) bunches. Lower fractional energies correspond to lower values of $\delta E/E$. The central beam energy is approximately 34~MeV and the measurement is performed at the X124 screen.}
  \label{fig:dpp}
\end{figure}

In addition to measuring the beam energy (as needed to compute the normalized-emittance values), sending the beam through the spectrometer beamline also provide a measurement of the energy spread by observing the beam at the X124 YAG screen at a point with vertical dispersion $\eta_y\simeq 0.29$~m. Figure~\ref{fig:dpp} presents the fraction-momentum distribution for the 1.6 and 3.2-nC bunches considered in the previous sections. The profile displays the characteristic asymmetrical energy distribution with a low-energy tail resulting from the RF-induced curvature in the longitudinal phase space; see Fig.\ref{fig:astraVSimpacttLPS}(a,b). We measured a full-width at half-maximum (FWHM) fractional energy spread of  $(9\pm 2)\times 10^{-4}$ and $(1.5 \pm 0.2)\times 10^{-3}$ for respectively 1.6 and 3.2-nC bunches. 

\section{Conclusion}
In summary, we have demonstrated the generation of high-charge (1.6 and 3.2~nC) magnetized bunches in a photoinjector. The measured eigenemittance partitions satisfy requirements for possible electron-cooling options foreseen in future electron-ion colliders. In practice, we anticipate that an injector optimized for electron cooling would deliver a lower 4D emittance (and hence lower $\varepsilon_-$ value) as the peak current nominally produced in the FAST photoinjector is approximately two orders of magnitude larger than required for electron cooling. Similarly, the measured total fractional energy spread is on the order of $10^{-3}$ (FWHM) while electron-cooling requires fractional spread in the range $[10^{-3}-10^{-4}]$ so that the longitudinal emittance attained in the considered L-band photoinjector is smaller than the one required for cooling. Our results should be understood as a "worst" case scenario as the configuration used in our experiment was not optimized to produce electron beams for cooling applications. Yet, the results confirm that photoinjectors can support the formation of high-charge magnetized beams with emittance partitions compatible with cooling requirements for future hadron accelerators. 
\section{Acknowledgements}
This work was support by the US Department of Energy, Office of Nuclear Physics, under contract DE-AC02-07CH11359 with Fermilab and DE-AC05-06OR23177 with Thomas Jefferson Laboratory. PP was partially supported by DOE grant DE-SC0018656 with Northern Illinois University. 
\bibliographystyle{elsarticle-num}
\bibliography{ref}

\end{document}